\documentstyle[12pt]{article}
\def \cn{Collaboration}

\newcommand{\thspace}{\kern.08333em}

\def \beq{\begin{equation}}
\def \eeq{\end{equation}}
\def \bea{\begin{eqnarray}}
\def \eea{\end{eqnarray}}

\def \Abar{\bar A}
\def \Dbar{\bar{D}^0}

\def \beq{\begin{equation}}
\def \eeq{\end{equation}}
\def \beqn{\begin{eqnarray}}
\def \eeqn{\end{eqnarray}}

\def \Dbar{\bar{D}^0}

\begin{document}

\rightline{CERN-TH/2002-331}
\rightline{hep-ph/0211282}
\rightline{November 2002}
\bigskip
\bigskip
\centerline{\bf IMPROVING BOUNDS ON $\gamma$ IN $B^{\pm} \to DK^{\pm}$ AND $B^{\pm,0} 
\to D X_s^{\pm,0}$}
\bigskip
\centerline{\it Michael Gronau\footnote{Permanent Address: Physics Department,
Technion -- Israel Institute of Technology, 32000 Haifa, Israel.}} 
\centerline{\it Theory Division, CERN} 
\centerline{\it CH--1211, Geneva 23, Switzerland}
\bigskip
\centerline{\bf ABSTRACT}
\vskip 2cm  

\begin{quote}
In view of recent experimental progress in rate and CP asymmetry measurements in
$B^{\pm} \to DK^{\pm}$, we reconsider information on the weak phase $\gamma$ 
which can be obtained from these processes. 
Model-independent inequalities are proven for $\sin^2\gamma$ in terms of two ratios of 
partial rates for $B^{\pm,0} \to D X_s^{\pm,0}$, where $X_s$ is any multiparticle charmless 
state carrying strangeness $\pm 1$. Good prospects are shown to exist for using 
these inequalities and CP asymmetry measurements in two body and multibody decays in order 
to improve present bounds on $\gamma$. 
\end{quote}

\vskip 3cm

\leftline{\qquad PACS codes:  12.15.Hh, 12.15.Ji, 13.25.Hw, 14.40.Nd}
\vfill
\newpage

The observation of CP violation in decays of $B$ mesons to $J/\psi$ and neutral 
kaons \cite{psiKS} is in good agreement with the prediction of the the Standard Model,
in which CP violation originates in a single phase $\gamma \equiv {\rm Arg} V^*_{ub}$
of the Cabibbo-Kobayashi-Maskawa (CKM) matrix. Further measurements of CP asymmetries 
in other $B$ decay processes are needed in order to establish the CKM hypothesis for 
CP violation on a firm ground, or to observe deviations from this simple picture.
So far CP violation in $B$ decays was observed only in processes involving $B^0-\bar B^0$
mixing, whereas the phase $\gamma$ has not yet been put to a direct test in $B$ decays.
It is therefore of great importance to search for {\it direct} CP violation in 
processes unaffected by uncertainties due to penguin amplitudes \cite{MG89}, where CP 
asymmetries have clean theoretical interpretations in terms of the weak phase $\gamma$.

One of the very early proposals for a clean measurement of $\gamma$   
is based on decays of the type $B^{\pm} \to D X_s^{\pm}$ \cite{GW}, where 
$X_s^{\pm}$ stands for a charged kaon or {\it any} few particle state with the same 
flavor quantum numbers as a charged kaon, e.g. $X_s = K, K^*, K\pi, K^*\pi$. The weak phase 
$\gamma$ occurs as the relative phase between two $B^-$ decay amplitudes into $D^0$ and 
$\Dbar$ flavor states, from $b \to c\bar u s$ and $b \to u \bar c s$, both contributing 
in decays to CP eigenstates, $D^0_{{\rm CP}\pm} = (D^0 \pm \Dbar)/\sqrt 2$.
In the original proposal all three $B^-$ decay amplitudes and a corresponding $B^+$ 
decay amplitude for a CP-eigenstate had to be measured in order to determine $\gamma$.
In the simplest case of two body decays, $X_s = K$, one of the four amplitudes, $A(B^- \to 
\bar D^0 K^-)$, is color-suppressed. Its measurement using hadronic $\Dbar$
decays is prohibited \cite{DU} due to interference with a comparable contribution from 
$B^- \to D^0 K^-$ followed by doubly-Cabibbo-suppressed (DCS) $D^0$ decays.
Nevertheless, it was noted in Ref.~\cite{MGr} that useful constraints on 
$\gamma$ can also be obtained without measuring this difficult mode. 
Several variants of this basic scheme were suggested, some of which rely on hitherto 
unmeasured and more difficult $B$ and $D$ decay modes \cite{var1}, and others which require 
extra assumptions about negligible rescattering effects \cite{var2}.

The magnitudes of all five amplitudes required for an implementation of this proposal,
$A(B^- \to D^0 K^-)$ and the four amplitudes $A(B^{\pm} \to D^0_{{\rm CP}\pm} K^{\pm})$, 
have already been measured.
The decay $B^- \to D^0 K^-$ and its charge-conjugate were observed several years 
ago \cite{BDK}. Recently branching ratios for the processes $D^0_{\rm CP} K^{\pm}$ 
were measured by the Belle collaboration \cite{Belle} both  for CP-even and odd states, 
and by the BABAR collaboration \cite{BABAR} for CP-even states.
CP asymmetry measurements in decays involving $D^0$ CP-eigenstates \cite{Belle,BABAR} are 
approaching a level for setting interesting bounds on the asymmetries. 
In addition, there exists new experimental information \cite{color} indicating that 
color-suppression of the ratio 
\beq
r\equiv |A(B^- \to \Dbar K^-)/A(B^- \to D^0 K^-)|
\eeq
is less effective than anticipated. This improves the feasibility of this method.

In view of these important developments, we wish to reconsider in this Letter the
implications which further improvements in these measurements will have on constraining 
$\gamma$. In particular, we make use of two inequalities \cite{MGr}
\beq\label{LIMIT}
\sin^2\gamma \leq R_{{\rm CP}\pm}~~,
\eeq
where we define for each of the two CP-eigenstates a ratio of charge-averaged rates   
\beq\label{Rcp}
R_{{\rm CP}\pm} \equiv \frac{2[\Gamma(B^- \to D^0_{{\rm CP}\pm} K^-) + \Gamma(B^+ \to 
D^0_{{\rm CP}\pm} K^+)]}{\Gamma(B^- \to D^0 K^-) + \Gamma(B^+ \to \Dbar K^+)}~~.
\eeq
We will find that, although these two constraints do not depend explicitly on $r$, and do 
not require a knowledge of $r$, in general they become stronger with increasing values of 
this parameter. For a reasonable estimate, $r\sim 0.2$, 
%MG added
one may encounter one of two possible situations: If the relevant final state interaction phase
$\delta$ is large, then one should soon measure for the first time direct CP violation in $B$ 
decays. On the other hand, if $\delta \le 30^{\circ}$, which can be verified by improving bounds 
on CP asymmetries, then the above constraints improve present bounds on $\gamma$. 

In the second part of the Letter we proceed to a general discussion of decays of the form 
$B^\pm \to D X_s^{\pm}$ and $B^0\,(\bar B^0) \to D X_s^0\,(\bar X_s^0)$, where $X_s^{\pm}$ 
and  $X_s^0\,(\bar X_s^0)$ are arbitrary charmless mutiparticle states with strangeness $\pm 1$. 
We will prove a generalization of Eq.~(\ref{LIMIT}) to multibody decays of this 
type. In the absence of color-suppression in most multibody decays, which implies larger values 
of corresponding $r$ parameters in these processes, these bounds are likely to provide 
stronger constraints on $\gamma$ than in the case of two body decays.
Our considerations apply to any multibody decay of this kind, for instance $B^\pm \to 
DK^\pm \pi^0$ and  the self-tagged $B^0\,(\bar B^0) \to D K^{\pm}\pi^{\mp}$, and are 
model-independent \cite{APS}.

Using notations for amplitudes as in \cite{GW,MGr} and disregarding a common strong phase,
\beq\label{A}
A(B^- \to D^0 K^-) = |A|~~,~~~~~A(B^- \to \Dbar K^-) = |\Abar| e^{i\delta} e^{-i\gamma}~~,
\eeq
we define in addition to the two ratios of charge-averaged rates (\ref{Rcp})
two pseudo asymmetries
\beq\label{Acp}
{\cal A}_{{\rm CP}\pm} \equiv \frac{2[\Gamma(B^- \to D^0_{{\rm CP}\pm} K^-) - \Gamma(B^+ \to 
D^0_{{\rm CP}\pm} K^+)]}{\Gamma(B^- \to D^0 K^-) + \Gamma(B^+ \to \Dbar K^+)}~~,
\eeq
from which ordinary CP asymmetries are obtained, $A_{{\rm CP}\pm} = {\cal A}_{{\rm CP}\pm}/
R_{{\rm CP}\pm}$.
Expressions of these measurables in terms of $r=|\Abar/A|,~\delta$ and $\gamma$ are readily 
obtained, neglecting tiny $D^0 - \Dbar$ mixing \cite{Silva} and using $D^0_{{\rm CP}\pm} = 
(D^0 \pm \Dbar)/\sqrt{2}$,
\beq\label{RCP}
R_{{\rm CP}\pm} = 1 + r^2 \pm 2r\cos\delta\cos\gamma~~,
\eeq
\beq\label{A+-}
{\cal A}_{{\rm CP}\pm}       = \pm 2r \sin\delta \sin\gamma~~,
\eeq 
where (\ref{RCP}) implies
\beq\label{S}
\frac{1}{2} (R_{{\rm CP}+} + R_{{\rm CP}-}) = 1 + r^2~~,
\eeq
and 
\beq\label{Rdif}
R_{{\rm CP}+} - R_{{\rm CP}-} = 4r\cos\delta\cos\gamma~~.
\eeq
The quantities $R_{{\rm CP}\pm}$ and ${\cal A}_{{\rm CP}\pm}$ hold 
information from which $r,~\delta$ and $\gamma$ can in principle be determined. The parameter
$r$ is given by (\ref{S}), and $\gamma$ is obtained up to a discrete ambiguity from 
$R_{{\rm CP}\pm}$ and ${\cal A}_{{\rm CP}\pm}$,
\beq\label{solution}
R_{{\rm CP}\pm} = 1 + r^2 \pm \sqrt{4r^2 \cos^2 \gamma - {\cal A}^2_{{\rm CP}\pm} \cot^2 \gamma}~~,
\eeq
where the $\pm$ signs on the right-hand-side correspond to even and odd CP states for
$\cos\delta\cos\gamma > 0$.

Plots of $R_{{\rm CP}\pm}$ as function of $\gamma$, for a few values of $r$ around 0.2 and 
asymmetries in the range of 
$0-30\%$, may be borrowed from \cite{KPI} plotting 
analogous quantities for the processes $B^0 \to K^+\pi^-$ and $B^+ \to K^0\pi^+$, which 
involve a similar algebra relating $\gamma$ to $B \to K\pi$ 
decay rates. 
Here one defines a ratio of charge-averaged decay rates, 
$R \equiv \Gamma(B \to K^{\pm}\pi^{\mp})/\Gamma(B^{\pm} \to K \pi^{\pm})$, which is given 
in terms of a pseudo asymmetry ($A_0$) in $B^0 \to K^+\pi^-$ and a ratio ($r$) of tree and
penguin amplitudes. In contrast to these decays, which involve a single ratio $R$, in
$B \to DK$ one measures two ratios for even and odd CP states. This resolves an ambiguity 
in the plots between $R > 1 + r^2$ and $R < 1 + r^2$ and allows for another measurable (\ref{Rdif}).
The plots of $R_{{\rm CP}\pm}$ as function of $\gamma$ can be used to study
the precision in $r,~R_{{\rm CP}\pm}$ and ${\cal A}_{{\rm CP}\pm}$ 
required to measure $\gamma$ to a given accuracy. In our case the accuracy is seen to improve 
with increasing values of $r$ due to a larger interference between $A(B^- \to D^0 K^-)$ and
$A(B^- \to \Dbar K^-)$. 

A crucial point is the actual value of $r$.  New experimental
information exists which relates to this value. Previously arguments based on na\"ive 
factorization \cite{BSW} seemed to imply 
that the amplitude $A(B^- \to \Dbar K^-)$ involves a suppression factor, $|a_2/a_1| = 0.25$ 
\cite{BS}, for the fact that the quark and antiquark making the kaon in $B^- \to \bar D^0 
K^-$ do not originate in the same weak current of the effective Hamiltonian describing 
$b \to u \bar c s$. This has led to a commonly accepted estimate 
$r \approx(|V_{ub}V^*_{cs}|/|V_{cb}V^*_{us}|)(|a_2/a_1|)\approx 0.1$.
Recent measurements \cite {color} of the color-suppressed process 
$\bar B^0 \to D^0\pi^0$ show, however, that color-suppression is less effective in 
this process than anticipated \cite{NP}, implying $a_2/a_1 \simeq 0.44$. Therefore, 
a more reasonable estimate is 
\beq\label{r2}
r \sim 0.2~~.
\eeq

As noted in the past \cite{MGr}, it is difficult to associate a theoretical 
uncertainty with this value. While the 
amplitude for $\bar B^0 \to D^0\pi^0$ involves a $b \to c$ transition with a heavy quark in 
the final state, $B^- \to \Dbar K^-$ follows from a $b \to u$ transition with a light
quark in the final state. The different flavor structure of the two operators and the 
different kinematics with which the heavy and light quarks emerge from the weak interaction 
imply different hadronic final state interaction effects in the two cases. This is expected 
to result in different color-suppression factors. 
Thus, while we will be using the value (\ref{r2}) as a guide for testing the sensitivity 
of this method, one should not exclude different values of $r$.

An important task of future studies is to determine $r$ experimentally without 
measuring $B^- \to \Dbar K^-$ \cite{other-r}. A useful lower bound on $r$
is obtained from Eq.~(\ref{Rdif}),
\beq
r \ge \frac{1}{4}\,|R_{{\rm CP}+} - R_{{\rm CP}-}|~~.
\eeq
If $r$ is as small as (\ref{r2}) it will be 
very difficult to determine its precise value from Eq.~(\ref{S}),
since the right-hand side is quadratic in $r$ and is expected to be 
only a few percent larger than one. Setting upper bounds on $r$ would also be useful.

%MG changed paragrpah
The information on $\delta$ and $\gamma$ obtained from ${\cal A}_{{\rm CP}\pm}$ and $R_{{\rm CP}\pm}$
is complementary to each other. While the asymmetries become larger 
for large values of $\sin\delta\sin\gamma$, both the deviation of $R_{{\rm CP}\pm}$ from one
and the difference $R_{{\rm CP}+} - R_{{\rm CP}-}$ increase with $\cos\delta\cos\gamma$. 
The two asymmetries (\ref{A+-}), which are equal in magnitude and opposite in sign, may 
be combined to yield an overall asymmetry, ${\cal A}_{{\rm CP}+} - {\cal A}_{{\rm CP}-}
= 4r \sin\delta \sin\gamma$.
Sizable CP asymmetries ${\cal A}_{{\rm CP}\pm}$ at a level of $20\%$, which could 
soon be observed \cite{Belle, BABAR}, require a large value of $|\sin\delta|$
corresponding to $|\delta| \ge 30^\circ$ (mod $\pi$).
A nonzero asymmetry would be an important observation by itself, demonstrating for the first 
time direct CP violation in $B$ decays. Observing nonzero CP asymmetries may 
however be difficult if $\delta$ is small. Currently there exists no information about 
$\delta$. A corresponding strong phase difference between isospin amplitudes 
in $B \to \bar D\pi$ decays was measured recently \cite{CLEOdelta} in the range $16^\circ - 
38^\circ$. QCD considerations suggest that this final state interaction phase occuring in 
$b \to c$ transitions is either perturbative or power suppressed in $1/m_b$ \cite{BBNS}. 
On the other hand, the phase $\delta$ is due to $b \to u$ transitions and may be different 
as mentioned above.

Anticipating that bounds on CP asymmetries will soon be improved to the level 
of $20\%$, thereby setting an upper bound on $|\delta|$, we will show that new constraints 
on $\gamma$ follow from $R_{{\rm CP}\pm}$ if $|\delta|$ is assumed to be smaller than about
$30^\circ$. This will be contrasted with weaker constraints in case that nonzero CP asymmetries 
are measured indicating larger values of $|\delta|$. 

Rewriting
\beq
R_{{\rm CP}\pm} = \sin^2\gamma + (r \pm \cos\delta\cos\gamma)^2 + 
\sin^2\delta\cos^2\gamma~~,
\eeq
one obtains the two simultaneous inequalities
\beq\label{LIM}
\sin^2\gamma \leq R_{{\rm CP}\pm}~~.
\eeq
These inequalities become useful when $R_{{\rm CP}\pm} < 1$ holds for either even or odd CP states.
This condition is fufilled in a major part of the $r,~\delta, \gamma$ parameter space 
because of the two opposite signs of the last term in (\ref{RCP}). The condition is equivalent 
to a rather weak requirement, $|\cos \delta \cos \gamma| > r/2$. Namely, Eqs.~(\ref{LIM}) imply 
nontrivial constraints on $\gamma$ when neither $\gamma$ nor $\delta$ lies too close to $\pi/2$. 
Since we are assuming this to be true for the strong phase $\delta$,
Eq.~(\ref{LIM}) provides useful bounds on $\gamma$ for values different from $\pi/2$. 
We note that, while the bounds (\ref{LIM}) themselves are not too useful when $\gamma$ is near 
$\pi/2$, such values of $\gamma$ can be tested and excluded by measuring $R_{{\rm CP}+} - 
R_{{\rm CP}-} = 4r\cos\delta\cos\gamma$.
 
% This is Table I
\begin{table}
\caption{Upper bounds on $\gamma$ (in degrees) obtained using Eqs.~(\ref{RCP}) and (\ref{LIM}).
Numbers in parentheses are corresponding maximal values of $R_{{\rm CP}\pm}$ for one of the two
CP-eigenstates.
\label{tab:bounds}}
\begin{center}
\begin{tabular}{c c c c c c} \hline \hline
%JR I suggest using \multicolumn for next line
% input value of $\gamma$ & upper bound & on $\gamma$ assuming & $|\delta| \le 30^{\circ}
Input value of $\gamma$ & \multicolumn{2}{c}{Upper bound on $\gamma$ assuming}
& $|\delta| \le 30^{\circ}$
%({\rm mod}~\pi$) 
& $\delta = 0$
& $\delta = 60$
%~({\rm mod}~\pi)$ 
\\ 
& $r = 0.1$ &  $r = 0.2$ & $r = 0.4$  & $r = 0.4$ & $r = 0.4$ \\
\hline
$50$   & $71~(0.90)$  &  $65~(0.82)$  &  $58~(0.71)$ & $53.5~(0.65)$ & $72~(0.90)$ \\
$60$   & $74~(0.92)$  &  $69~(0.87)$  &  $64~(0.81)$ & $60.7~(0.76)$ & $78~(0.96)$ \\
$70$   & $77~(0.95)$  &  $74~(0.92)$  &  $74~(0.92)$ & $70.3~(0.89)$ & $-~(1.02)$ \\
$80$   & $82~(0.98)$  &  $82~(0.98)$  &  $-~~(1.04)$  & $-~~~(1.02)$ & $-~(1.09)$ \\ \hline \hline
\end{tabular}
\end{center}
\end{table}

The bounds on $\gamma$ depend on the value of $r$.
Since we are assuming the CKM framework, we disregard a discrete ambiguity in $\gamma$, 
taking its value to be smaller than $\pi/2$ \cite{BPS}. For illustration, we calculate
in Table 1 upper bounds on $\gamma$ obtained from Eqs.~(\ref{RCP}) and (\ref{LIM}) for 
three values of $r$, $r=0.1,~0.2,~0.4$. The value $r = 0.4$, which may be an overestimate
for the case of two body decays, is a realistic value for multibody decays which we discuss
below. We include it in the present discussion for a later reference. Values of $R_{{\rm CP}\pm}$
which depend on $\gamma$ and resulting bounds on this phase are computed for input values in the 
range $50^{\circ} \le \gamma \le 80^{\circ}$ permitted by CKM fits \cite{BPS}. 
In computing the bounds for $r=0.1,~0.2,~0.4$ in the second, third and fourth columns we 
assume $|\delta| \le 30^\circ~({\rm mod}~\pi)$ which can be verified by CP asymmetry measurements. 
The bounds in the fifth and sixth columns correspond to $\delta =0$ and $\delta = 60^\circ$, 
respectively. Also given in parentheses in the second, third and fourth columns are maximal 
values of $R_{{\rm CP}\pm}$ for one of the two CP-eigenstates, which are obtained 
for $\delta_{\rm max} = 30^{\circ}$ (mod $\pi$). Smaller 
values of $R_{{\rm CP}\pm}$, and corresponding stronger upper bounds on $\gamma$, are obtained for 
$\delta =0$ as shown in the fifth column. On the other hand, the bounds become weaker 
when $\delta$ is large, as demonstrated for $\delta = 60^\circ$ in the last column.
In this case one should soon observe a CP asymmetry in $B^{\pm} \to D^0_{{\rm CP}\pm} K^{\pm}$.
 
We note that as $r$ increases the bounds on $\gamma$ become stronger. For $r=0.2$, as estimated in 
(\ref{r2}), and assuming $|\delta| \le 30^\circ~({\rm mod}~\pi)$, the upper limits on $\gamma$ 
already provide useful 
information on the weak phase beyond CKM fits. For instance, for $\gamma = 50^{\circ}$ the deviation 
of the lower $R_{{\rm CP}\pm}$ value from one is substantial, implying $\gamma \le 65^{\circ}$. 
The corresponding difference 
$|R_{{\rm CP}+} - R_{{\rm CP}-}| = 0.45$ is quite large. For $r=0.4$ 
the limits are rather close to the input values of $\gamma$, in particular for $\delta = 
0~({\rm mod}~\pi)$. If the actual value of $\gamma$ is near $50^{\circ}$ then the bound fixes the 
weak phase to a very narrow range of several degrees. In this case the values of 
%MG added | |
$|R_{{\rm CP}+} - R_{{\rm CP}-}|$ are $0.89$ and $1.03$, corresponding to $\delta = 30^{\circ}$ (mod $\pi$) 
and $\delta = 0~({\rm mod}~\pi)$, respectively.
One notes that, while bounds on CP asymmetries do not distinguish between the two possibilities that
$\delta$ is near zero or near $\pi$, the former case seems to be favored by theory \cite{BBNS}. 
Therefore, one expects $R_{{\rm CP}+} > R_{{\rm CP}-}$.
We conclude that, although $r$ may not be determined accurately experimentally, there exist 
good prospects, in terms of reasonable values of $r$ and $\delta$ which is assumed not to be too
large, for improving present bounds on $\gamma$ using measured values of $R_{{\rm CP}\pm}$.

The small value of $r\equiv |A(B^- \to \Dbar K^-)/A(B^- \to D^0 K^-)|$ in (\ref{r2}) follows 
from color-suppression in the two body decay $B^- \to \Dbar K^-$ on top of a modest 
CKM-suppression. Color suppression arguments are unknown to hold in practice in multibody decays, and
furthermore do not apply formally to most multibody decays of the type $B^-\,(\bar B^0) \to \Dbar 
X_s^{-,0}$ \cite{exception}. For instance, the processes $B^- \to \Dbar K^- \pi^0, B^- \to \Dbar 
K^- \pi^+\pi^-$ and the self-tagged $\bar B^0 \to \Dbar K^-\pi^+$ involve the same color factors as 
the coresponding processes with $D^0$ in the final state.
Thus, disregarding a possible suppression due to form factors or other dynamical factors,  
one expects the corresponding ratio of $\Dbar$ and $D^0$ amplitudes in most multibody $B \to D X_s$
decays to be larger than in two body decays. As we saw now, the two inequalities
(\ref{LIM}) become stronger as $r$ increases. An interesting question is therefore 
whether inequalities similar to (\ref{LIM}) hold also in multibody decays. If this were the case, 
then one would be able to apply such inequalities to these decays in order to obtain stronger 
constraints on $\gamma$. In the remaining part of this Letter we will prove that such 
inequalities hold in general and we will study their consequences.
 
In a multibody decay of the type $B^- \to D X_s^-$, and in a similar neutral $B$ decay $\bar B^0 
\to D X_s^0$, one may generalize Eqs.~(\ref{A}) 
to hold at any point $p$ in the multibody phase space,
\beq\label{Ap}
A(B^- \to (D^0 X_s^-)_p) = A_p~~,~~~~~A(B^- \to (\Dbar X_s^-)_p) = \Abar_p e^{-i\gamma}~~,
\eeq
and consequently
\beq\label{ACPp}
A(B^- \to (D^0_{{\rm CP}\pm} X_s^-)_p) = \frac{1}{\sqrt 2}(A_p \pm \Abar_p e^{-i\gamma})~~,
\eeq
where $A_p$ and $\Abar_p$ are complex amplitudes involving final state interaction phases
which depend on the point $p$ in phase space. For instance, in $B^- \to D K^-\pi^0$~~$p$
is a point in a Dalitz plot and the magnitudes and complex phases of $A_p$ and $\Abar_p$, 
which depend on resonance structures in the two channels, vary from one point to another. 
This seems to pose a serious problem in applying the method \cite{GW} or any of its variants 
to multibody decays in order to determine $\gamma$. Any such attempt would be strongly 
model-dependent, since it requires modeling the amplitudes $A_p$ and $\Abar_p$ as functions 
of $p$ in terms of assumed resonance structures in the two channels. For a very recent attempt, 
see Ref.~\cite{APS}. Our following considerations are, however, model-independent.

Let us consider partial rates for the four processes in Eqs.~(\ref{Ap}) and (\ref{ACPp}),
\beq
\Gamma(B^- \to D^0 X_s^-) = \int dp\,|A_p|^2~~,~~~~~\Gamma(B^- \to \Dbar X_s^-) = \int dp 
\,|\Abar_p|^2~~,
\eeq
\beq\label{GammaCP}
\Gamma(B^- \to D^0_{{\rm CP}\pm} X_s^-) = \frac{1}{2}\left (\int dp\,|A_p|^2 + \int dp\,|\Abar_p|^2
\right ) \pm \int dp\,{\rm Re}(A_p\Abar^*_p e^{i\gamma})~~,
\eeq
where integration over phase space may be either complete or partial.
The corresponding $B^+$ decay rates for decays to CP-eigenstates are obtained by changing the 
sign of $\gamma$ in Eq.~(\ref{GammaCP}).
Defining ratios of partial rates and pseudo asymmetries,
\bea\label{rs}
r_s^2 & \equiv & \frac{\Gamma(B^- \to \Dbar X_s^-)}{\Gamma(B^- \to D^0 X_s^-)}~~,\\
R_{{\rm CP}\pm}(X_s) & \equiv & \frac{2[\Gamma(B^- \to D^0_{{\rm CP}\pm} X_s^-) + \Gamma(B^+ \to 
D^0_{{\rm CP}\pm} X_s^+)]}{\Gamma(B^- \to D^0 X_s^-) + \Gamma(B^+ \to \Dbar X_s^+)}~~,\\
{\cal A}_{{\rm CP}\pm}(X_s) & \equiv & \frac{2[\Gamma(B^- \to D^0_{{\rm CP}\pm} X_s^-) - \Gamma(B^+ 
\to D^0_{{\rm CP}\pm} X_s^+)]}{\Gamma(B^- \to D^0 X_s^-) + \Gamma(B^+ \to \Dbar X_s^+)}~~,
\eea
we find
\bea\label{Rs}
R_{{\rm CP}\pm}(X_s) & = & 1 + r_s^2 \pm 2\cos\gamma\,\frac{{\rm Re}(\int dp\,A_p\Abar^*_p)}
{\int dp\,|A_p|^2}~~,\\
{\cal A}_{{\rm CP}\pm}(X_s) & = & \pm 2\sin\gamma\frac{{\rm Im}(\int dp\,A_p\Abar^*_p)}
{\int dp\,|A_p|^2}~~.
\eea

Denoting
\beq\label{kappa}
\kappa e^{i\delta_s} \equiv \frac{\int dp\,A_p \Abar^*_p}{\sqrt{\int dp\,|A_p|^2\,\int dp\,
|\Abar_p|^2}}~~,
\eeq
where the Schwarz inequality for the two complex vectors $A_p$ and $\Abar_p$ implies 
$0 \le \kappa \le 1$, one obtains
\bea\label{Rs}
R_{{\rm CP}\pm}(X_s) & = & 1 + r_s^2 \pm 2\kappa r_s\cos\delta_s\cos\gamma~~,\\
\label{As}
{\cal A}_{{\rm CP}\pm}(X_s) & = & \pm 2\kappa r_s \sin\delta_s\sin\gamma~~.
\eea
Eq.~(\ref{Rs}) leads immediately to
\beq\label{LIMs}
\sin^2\gamma \le R_{{\rm CP}\pm}(X_s)~~.
\eeq
The expressions of $R_{{\rm CP}\pm}(X_s)$ and ${\cal A}_{{\rm CP}\pm}(X_s)$ become identical to those 
of two body decays for $\kappa = 1$, namely 
when $A_p$ and $\Abar_p$ are parallel (i.e. proportional) to each other. This is the case in which
rate and asymmetry measurements are most sensitive to the weak phase $\gamma$.
An extreme and unfavorable case, $\kappa = 0$, occurs when $A_p$ and $\Abar_p$ are orthogonal to 
each other. An upper bound on $\kappa$, which is saturated when relative phases between 
$A_p$ and $\bar A_p$ vanish  and which becomes weak when these phases are large, can be expressed in 
terms of measurable differential rates,
\beq\label{boundkappa}
\kappa \le \frac{\int dp\,|A_p||\Abar_p|}{\sqrt{\int dp\,|A_p|^2\,\int dp\,|\Abar_p|^2}}~~.
\eeq

Eqs.~(\ref{Rs}), (\ref{As}) and the bound (\ref{LIMs}) are quite powerful. They apply to any 
multibody decay of the type under discussion and to an arbitrary choice of phase space over which one 
integrates. Their advantage over the corresponding relations in two body decays is threefold:
\begin{enumerate}
\item Multibody decays are expected to have larger branching ratios than two body decays.
\item Since most multibody decays of the type $B^- \to \Dbar X_s^-$ and $\bar B^0 \to \Dbar X_s^0$ 
are not color-suppressed, their measurements using hadronic $\Dbar$ decays are less affected by 
interference with doubly Cabibbo-suppressed $D^0$ decays in $B^-\to D^0 X_s^-$ and $\bar B^0 \to 
D^0 X_s^0$, respectively. This allows a reasonably accurate direct measurement of $r_s$ in these 
processes. 
\item In general the parameter $r_s$ is expected to be larger than $r$ which is 
color-suppressed. A  typical estimate, based only on CKM factors, is $r_s 
\approx|V_{ub}V^*_{cs}|/|V_{cb}V^*_{us}|$ $\approx 0.4$.  
One expects that in certain decays and in specific regions of phase space the value 
of $r_s$ may be larger than $0.4$ due to a dynamical enhancement of $\Abar_p$ relative to $A_p$.
The sensitivity to $\gamma$ grows with $\kappa r_s$, where there exists no direct measurement
for $\kappa$ except the upper bound (\ref{boundkappa}). One may choose judiciously, or by scanning 
over different regions of phase space, 
regions which minimize the lower of the two $R_{{\rm CP}\pm}(X_s)$ values. This would correspond to
maximizing the value of $\kappa r_s$ while keeping the phase $\delta_s$ as small as possible. 
On the other hand, large CP asymmetries correspond to large values of $\delta_s$. 
For instance, in $B \to D K\pi$~ the phase $\delta_s$ is expected to become small as one moves 
away from $K$ and $D(D_s)$ resonance states, and to increase as one approaches 
the resonance bands \cite{AEGS,APS}. Upper bounds on $\gamma$ were calculated in the fourth and fifth 
columns of Table 1 for $\kappa=1, r_s = 0.4, \delta_s \le 30^\circ$ and $\delta_s =0$, and were shown 
to be considerably stronger than for $r=0.2$. They may correspond approximately to realistic situations 
in multibody decays. The strongest bounds on $\gamma$ are obtained when applying 
$R_{{\rm CP}\pm}(X_s)$ and $A_{{\rm CP}\pm}(X_s)$ to regions of phase space in which $A_p$ and 
$\Abar_p$ are proportional to each other, corresponding to $\kappa = 1$. In this case an algebraic 
solution for $\gamma$ can be obtained as shown in Eq.~(\ref{solution}).
\end{enumerate}

Before concluding, let us comment on the measurement of $R_{{\rm CP}\pm}$ upon which the 
proposed bounds on $\gamma$ depend. The ratios $R_{{\rm CP}\pm}$ involve $B$ decay rates into 
CP and flavor states of neutral $D$ mesons. They are measured by observing $D^0$ decay modes 
involving even CP ($K^+K^-,~\pi^+\pi^-$), odd CP ($K_S\pi^0,~K_S\phi,~K_S\omega,~K_S\rho, 
K_S\eta,~K_S\eta'$), and flavor states ($K^-\pi^+,~K^-\pi^+\pi^0,~K^-\pi^+\pi^+\pi^-,
K^-\pi^+\pi^+\pi^-\pi^0$). It may seem that accurate measurements of $R_{{\rm CP}\pm}$ 
require precise knowledge of $D^0$ decay branching ratios into these states, which is the current
situation in some decay modes but not in all. There exists, however, a way in which $R_{{\rm CP}\pm}$
may be measured independent of $D^0$ decay branching ratios. 
Let us define two ratios which do not depend on $D^0$ decay branching ratios,
\beq\label{Kpi}
R(K/\pi) \equiv \frac{{\cal B}(B^- \to D^0 K^-)}{{\cal B}(B^- \to D^0 \pi^-)}~~,
\eeq
\beq\label{KpiCP}
R(K/\pi)_{{\rm CP}\pm} \equiv \frac{{\cal B}(B^- \to D^0_{{\rm CP}\pm} K^-) + 
{\cal B}(B^+ \to D^0_{{\rm CP}\pm} K^+)}{{\cal B}(B^- \to D^0_{{\rm CP}\pm} \pi^-) + 
{\cal B}(B^+ \to D^0_{{\rm CP}\pm} \pi^+)}~~.
\eeq
Using 
\beq
A(B^+ \to D^0_{{\rm CP}\pm}\pi^+) \approx A(B^- \to D^0_{{\rm CP}\pm}\pi^-) \approx 
\frac{1}{\sqrt 2}\,A(B^- \to D^0\pi^-)~~,
\eeq
where one neglects a term $(|V_{ub}V^*_{cd}/V_{cb}V^*_{ud}|)(|a_2/a_1|) = r|V_{us}V_{cd}/V_{ud}V_{cs}| 
\approx 0.01$, one finds
\beq
R_{{\rm CP}\pm} = \frac{R(K/\pi)_{{\rm CP}\pm}}{R(K/\pi)}~~.
\eeq

The ratios (\ref{Kpi}) and (\ref{KpiCP}) were measured in \cite{BDK,Belle,BABAR}.
While the current average value, $R(K/\pi) = 0.0819 \pm 0.0037$, involves only a small error,
errors are still large in the two measurements, $R(K/\pi)_{{\rm CP}+} = 0.125 \pm 0.036 \pm 0.010$ 
\cite{Belle} and $0.074 \pm 0.017 \pm 0.006$ \cite{BABAR}, as well as in the single measurement, 
$R(K/\pi)_{{\rm CP}-} = 0.119 \pm 0.028 \pm 0.006$ \cite{Belle}.
The implied averages, $R_{{\rm CP}+} = 1.15 \pm 0.22$ and $R_{{\rm CP}-} = 1.45 \pm 0.36$,
are still consistent with $R_{{\rm CP}+} - R_{{\rm CP}-} =4r\cos\delta\cos\gamma = 0$.
This situation should change when errors are reduced.
As we have argued, it is unlikely that both $R_{{\rm CP}+}$ and $R_{{\rm CP}-}$ are larger than one.
Since $\cos\delta > 0$ seems to be favored theoretically, one would expect $R_{{\rm CP}-}$ 
to be smaller than one and to provide new bounds on $\gamma$. This requires some reduction of the 
errors in $R(K/\pi)_{{\rm CP}\pm}$.

In conclusion, we have shown that several two body decays $B^{\pm} \to D K^{\pm}$, for 
which data exist, have the potential of improving present bounds on the weak phase $\gamma$,
in particular if CP asymmetries are not soon observed in decays to $D^0$ CP-eigenstates..
We argued that multibody decays of this class, for both charged and neutral $B$ mesons, are expected 
to even do better. 
Measuring  $R_{{\rm CP}\pm}(X_s) = 0.60 \pm 0.05$ for one of the two CP-eigenstates in any of these 
decays would determine $\gamma$ to within several degrees, $\gamma = 51^{\circ} \pm 3^{\circ}$, 
approaching the present level of precision in $\beta$ \cite{BPS}. On the other hand, a measurement
$R_{{\rm CP}\pm}(X_s) < 0.5$, corresponding to $\gamma < 45^{\circ}$, would 
be a signature for physics beyond the Standard Model.  

\bigskip
%\bigskip
%\centerline{\bf  ACKNOWLEDGEMENTS}
%\bigskip

I am grateful to the CERN Theory Division for its kind hospitality, and I wish to thank 
J. L. Rosner for useful comments.
\bigskip

% Journal and other miscellaneous abbreviations for references
% Phys. Rev. D style
\def \ajp#1#2#3{Am.~J.~Phys.~{\bf#1}, #2 (#3)}
\def \apny#1#2#3{Ann.~Phys.~(N.Y.) {\bf#1}, #2 (#3)}
\def \app#1#2#3{Acta Phys.~Polonica {\bf#1}, #2 (#3)}
\def \arnps#1#2#3{Ann.~Rev.~Nucl.~Part.~Sci.~{\bf#1}, #2 (#3)}
\def \cmp#1#2#3{Commun.~Math.~Phys.~{\bf#1}, #2 (#3)}
\def \cn{Collaboration}
\def \ib{{\it ibid.}~}
\def \ibj#1#2#3{~{\bf#1}, #2 (#3)}
\def \ijmpa#1#2#3{Int.~J.~Mod.~Phys.~A {\bf#1}, #2 (#3)}
\def \ite{{\it et al.}}
\def \jhep#1#2#3{JHEP~{\bf#1}, #2 (#3)}
\def \jmp#1#2#3{J.~Math.~Phys.~{\bf#1}, #2 (#3)}
\def \jpg#1#2#3{J.~Phys.~G {\bf#1}, #2 (#3)}
\def \mpla#1#2#3{Mod.~Phys.~Lett.~A {\bf#1}, #2 (#3)}\def \ib{{\it ibid.}~}
\def \ibj#1#2#3{~{\bf#1}, #2 (#3)}
\def \ijmpa#1#2#3{Int.~J.~Mod.~Phys.~A {\bf#1}, #2 (#3)}
\def \ite{{\it et al.}}
\def \jmp#1#2#3{J.~Math.~Phys.~{\bf#1}, #2 (#3)}
\def \jpg#1#2#3{J.~Phys.~G {\bf#1}, #2 (#3)}
\def \mpla#1#2#3{Mod.~Phys.~Lett.~A {\bf#1}, #2 (#3)}
\def \nc#1#2#3{Nuovo Cim.~{\bf#1}, #2 (#3)}
\def \npb#1#2#3{Nucl.~Phys. B~{\bf#1}, #2 (#3)}
\def \pisma#1#2#3#4{Pis'ma Zh.~Eksp.~Teor.~Fiz.~{\bf#1}, #2 (#3) [JETP
Lett. {\bf#1}, #4 (#3)]}
\def \pl#1#2#3{Phys.~Lett.~{\bf#1}, #2 (#3)}
\def \plb#1#2#3{Phys.~Lett.~B {\bf#1}, #2 (#3)}
\def \pr#1#2#3{Phys.~Rev.~{\bf#1}, #2 (#3)}
\def \pra#1#2#3{Phys.~Rev.~A {\bf#1}, #2 (#3)}
\def \prd#1#2#3{Phys.~Rev.~D {\bf#1}, #2 (#3)}
\def \prl#1#2#3{Phys.~Rev.~Lett.~{\bf#1}, #2 (#3)}
\def \prp#1#2#3{Phys.~Rep.~{\bf#1}, #2 (#3)}
\def \ptp#1#2#3{Prog.~Theor.~Phys.~{\bf#1}, #2 (#3)}
\def \rmp#1#2#3{Rev.~Mod.~Phys.~{\bf#1}, #2 (#3)}
\def \rp#1{~~~~~\ldots\ldots{\rm rp~}{#1}~~~~~}
\def \stone{{\it B Decays}, edited by S. Stone (World Scientific,
Singapore, 1994)}
\def \yaf#1#2#3#4{Yad.~Fiz.~{\bf#1}, #2 (#3) [Sov.~J.~Nucl.~Phys.~{\bf #1},
#4 (#3)]}
\def \zhetf#1#2#3#4#5#6{Zh.~Eksp.~Teor.~Fiz.~{\bf #1}, #2 (#3) [Sov.~Phys.
- JETP {\bf #4}, #5 (#6)]}
\def \zpc#1#2#3{Zeit.~Phys.~C {\bf#1}, #2 (#3)}


\begin{thebibliography}{99}
\bibitem{exception} Color factors in multibody decays are inferred formally from quark 
diagrams in which a light $q\bar q$ pair implies an extra factor of $N_c$. In the
special case of $B^- \to \Dbar K_S \pi^-$ the amplitude obtains only a color-suppressed 
contribution.
\end{thebibliography}
\end{document}